\documentclass[twocolumn,prl,showpacs,preprintnumbers,amsmath,amssymb]{revtex4}
\usepackage{graphicx}
\usepackage{dcolumn}
\usepackage{bm}

\begin{document}

\title{Electron Spin Dynamics in Impure Quantum Wells for Arbitrary Spin-Orbit Coupling}
\author{C. Grimaldi}\email{claudio.grimaldi@epfl.ch}
\affiliation{Ecole Polytechnique F\'ed\'erale de Lausanne, LPM,
Station 17, CH-1015 Lausanne, Switzerland}

\begin{abstract}

Strong interest has arisen recently on low-dimensional systems
with strong spin-orbit interaction due to their enhanced
efficiency in some spintronic applications. Here, the time
evolution of the electron spin polarization of a disordered
two-dimensional electron gas is calculated exactly within the
Boltzmann formalism for arbitrary couplings to a Rashba spin-orbit
field. It is shown that the classical Dyakonov-Perel mechanism of
spin relaxation gets deeply modified for sufficiently strong
Rashba fields, in which case new regimes of spin decay are
identified. These results suggest that spin manipulation can be
greatly improved in strong spin-orbit interaction materials.
\end{abstract}


\maketitle

The physics of transport of electron spins in low-dimensional
systems has become a popular theme of research due to the possible
impact in future electronic applications\cite{prinz,fabian}. Key
subjects of studies concern the problem of controlling the
electron spin polarization through the tailoring of the
spin-orbit (SO) interaction \cite{dattadas,loss}, and the
knowledge of the physical parameters governing the spin
relaxation time $\tau_s$. Main sources of SO coupling are the
Rashba interaction arising from structural inversion asymmetries
of low-dimensional structures such as quantum wells or
wires\cite{rashba}, and the Dresselhaus interaction present in
bulk crystals lacking symmetry of inversion\cite{dressel}. In the
presence of Rashba and/or Dresselhaus interactions, $\tau_s$
basically arises from the randomization of the SO
(pseudo)magnetic field induced by momentum scattering with
impurities or other interactions (Dyakonov-Perel mechanism
\cite{DP}).

Of great interest for spintronic applications are materials with
strong SO interaction since they are more effective
spin-manipulators \cite{fabian} and spin-current generators
through the spin-Hall effect \cite{sinova}. In this respect,
systems of interest may be, for example, HgTe quantum wells whose
Rashba interaction leads to SO band splitting $\Delta_{\rm so}$
as large as  $\sim 30$ meV\cite{gui}, or surface states of metals
and semimetals which display giant SO band splittings of the
order of 0.2 eV or surface states of metals and semimetals which
display giant SO band splittings of the order of 0.2 eV or larger
\cite{rote1,koroteev}, or even the heavy Fermion superconductor
CePt$_3$Si \cite{bauer} whose lack of bulk inversion symmetry
leads to $\Delta_{\rm so} \simeq 50-200$ meV \cite{samo1}. For
all these systems, the SO splitting $\Delta_{\rm so}$ is no
longer the smallest energy scale, as in most semiconductors, and
can be comparable to the Fermi energy $\epsilon_F$ or even larger.

Despite that the time evolution of the spin polarization ${\bf
S}$ has been thoroughly studied in the weak SO limit $\Delta_{\rm
so}/\epsilon_F = 0$ \cite{DP,wu,glazov,saxena}, only few partial
studies have been devoted to the strong SO case $\Delta_{\rm
so}/\epsilon_F\neq 0$ \cite{ivchenko,burkov,lechner}. Of
particular interest is the problem of assessing the role of
$\Delta_{\rm so}/\epsilon_F$ on the spin relaxation mechanism and
of clarifying to which extent the classical Dyakonov-Perel (DP)
description \cite{DP} gets modified by finite values of
$\Delta_{\rm so}/\epsilon_F$.

In this letter, a generalized kinetic equation for ${\bf S}$ is
formulated for arbitrary strengths of the SO interaction and in
the presence of spin-conserving coupling with impurities. For the
case of a two-dimensional electron gas confined in the $x-y$
plane subjected to the Rashba interaction, the kinetic equation
for the $z$-component of ${\bf S}$ is solved explicitly for any
value of $\Delta_{\rm so}/\epsilon_F$. It is found that for a
sufficiently strong Rashba interaction the DP relaxation
mechanism gets modified, with the spin polarization displaying a
slow (power law) decay with time. Furthermore, in the extreme
$\Delta_{\rm so}/\epsilon_F \gg 1$ limit, a fast exponential
decay is obtained with $\tau_s$ proportional to the momentum
scattering time $\tau_{\rm p}$, {\it i.e.}, spin relaxation is
enhanced by disorder. These findings suggest that tailoring of
spin memory through $\Delta_{\rm so}/\epsilon_F$ can be much more
effective than previously thought.

\begin{figure}[t]
\protect
\includegraphics[scale=0.32]{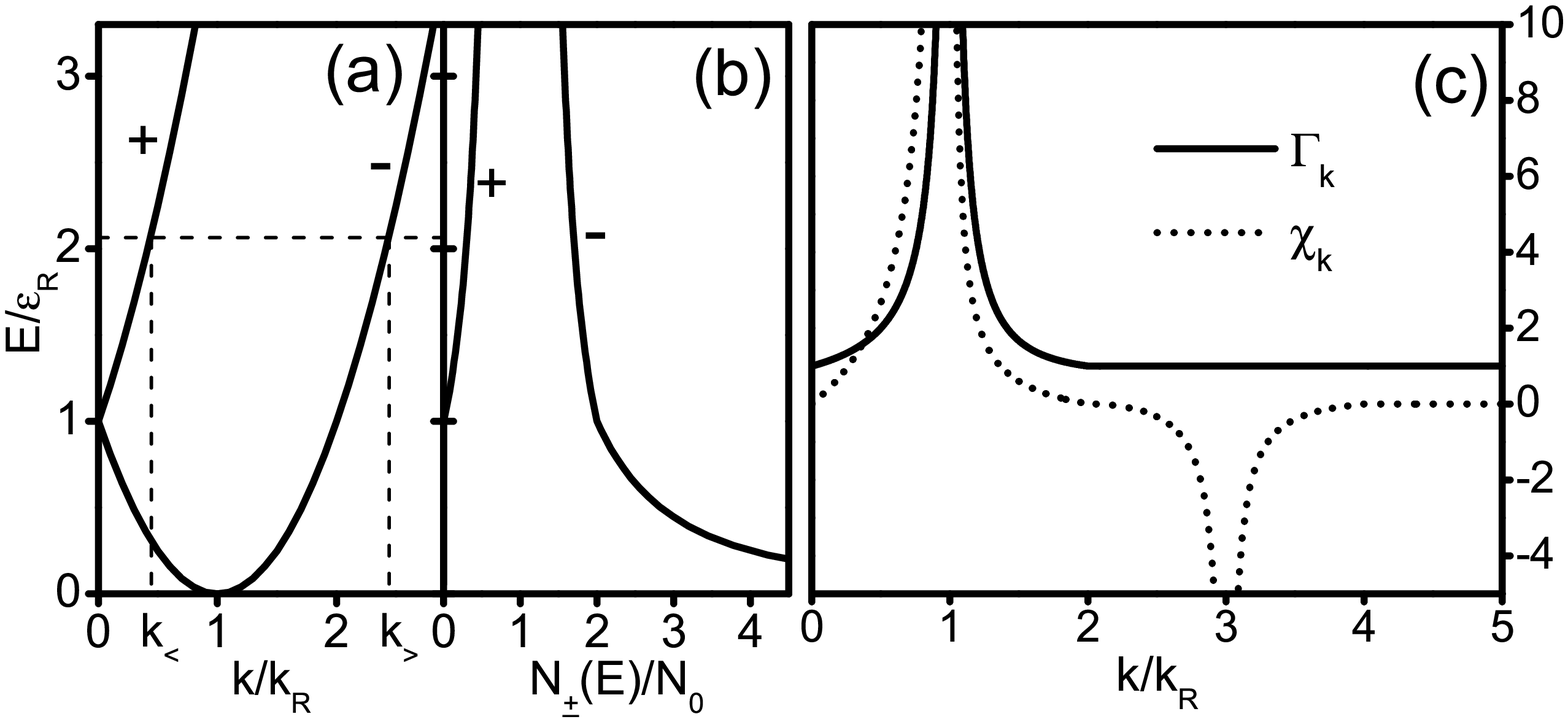}
\caption{(a): Rashba SO split electron dispersions
$\frac{\hbar^2}{2m}(k\pm k_R)^2$ in units of
$\epsilon_R=\frac{m}{8}\gamma_R^2$. The horizontal dashed line
indicates the Fermi level. (b): density of states for the $\pm$
bands. (c): plots of $\Gamma_k$, Eq.(\ref{gamma3}), and $\chi_k$,
Eq.(\ref{chi3}). } \label{fig1}
\end{figure}

Let us consider an electron gas whose non-interacting hamiltonian
$H_0$ is:
\begin{equation}
\label{ham1} H_0=\sum_{{\bf k},s}\epsilon_{{\bf
k}}c^\dagger_{{\bf k}s}c_{{\bf k}s}+\frac{\hbar}{2}\sum_{{\bf
k},ss'}\mathbf{\Omega}_{\bf
k}\!\cdot\!\widehat{\mbox{\boldmath$\sigma$}}_{ss'}\,c^\dagger_{{\bf
k}s}c_{{\bf k}s'},
\end{equation}
where $c^\dagger_{{\bf k}s}$ ($c_{{\bf k}s}$) is the creation
(annihilation) operator for an electron with momentum ${\bf k}$
and spin index $s=\uparrow,\downarrow$,
$\widehat{\mbox{\boldmath$\sigma$}}$ is the spin-vector operator
with components
($\widehat{\sigma}^x,\widehat{\sigma}^y,\widehat{\sigma}^z$) given
by the Pauli matrices, and $\mathbf{\Omega}_{\bf k}$ is a ${\bf
k}$ dependent SO pseudopotential vector whose explicit form is
not essential for the moment. In the above expression,
$\epsilon_{{\bf k}}$ is the electron dispersion in the absence of
SO coupling. Let us consider as momentum-relaxation mechanism a
coupling $V_{{\bf k}{\bf k}'}$ to spin-conserving impurities
located at random positions ${\bf R}_i$. Additional interaction
channels as those provided by phonons or electron-electron
couplings can be treated as well by following the same steps
described below. Assuming that $V_{{\bf k}{\bf k}'}$ is switched
on at time $t=t_0$, the time evolution of ${\bf S}(t)$ for
$t>t_0$ can be obtained from the equation of motion of the
density matrix $\rho(t)$\cite{blum}:
\begin{equation} \label{dm5}
\frac{d\widehat{\rho}_{{\bf k}{\bf
k}'}(t)}{dt}=-\frac{i}{\hbar}[\widehat{E}_{\bf
k}\widehat{\rho}_{{\bf k}{\bf k}'}(t)-\widehat{\rho}_{{\bf k}{\bf
k}'}(t)\widehat{E}_{{\bf
k}'}]-\frac{i}{\hbar}\widehat{\Gamma}_{{\bf k}{\bf k}'}(t),
\end{equation}
where $\widehat{\rho}_{{\bf k}{\bf k}'}(t)$ is a $2\times 2$
matrix with elements $\rho_{{\bf k}s,{\bf k}'s'}(t)=\langle{\bf
k}s|\rho(t)|{\bf k}'s'\rangle$, $\widehat{E}_{\bf k}=\epsilon_{\bf
k}-\mu+\frac{\hbar}{2}\mathbf{\Omega}_{\bf
k}\!\cdot\!\widehat{\mbox{\boldmath$\sigma$}}$, and
\begin{eqnarray}
\label{dm7} \widehat{\Gamma}_{{\bf k}{\bf k}'}(t)&=&\sum_{{\bf
p}i}\left[V_{{\bf k}{\bf p}}e^{i{\bf R}_i\cdot({\bf k}-{\bf
p})}\widehat{\rho}_{{\bf p}{\bf k}'}(t)\right.\nonumber \\
& &\left.- V_{{\bf p}{\bf k}'} e^{i{\bf R}_i\cdot({\bf p}-{\bf
k}')}\widehat{\rho}_{{\bf k}{\bf p}}(t)\right].
\end{eqnarray}
Equation (\ref{dm5}) can be formally integrated and, after the
usual adiabatic ($t_0\rightarrow -\infty$) and Markov
approximations \cite{blum,rossi}, $\widehat{\rho}_{{\bf k}{\bf
k}'}(t)$ can be expressed as:
\begin{equation}
\label{dm8} \widehat{\rho}_{{\bf k}{\bf k}'}(t)\simeq
-\frac{i}{\hbar}\int_0^\infty \!\!dt'e^{-\delta t'}
e^{-i\widehat{E}_{\bf k}t'/\hbar}\widehat{\Gamma}_{{\bf k}{\bf
k}'}(t)e^{i\widehat{E}_{{\bf k}'}t'/\hbar}
\end{equation}
where the limit $\delta\rightarrow 0^+$ must be taken after the
integration. By using the anticommutation property of the Pauli
matrices,
$\widehat{\sigma}_i\widehat{\sigma}_j+\widehat{\sigma}_j\widehat{\sigma}_i=
\widehat{1}\delta_{ij}$, the exponential operators in
Eq.(\ref{dm8}) can be put in a form more suitable for integration
over $t'$:
\begin{equation}
\label{dm16} e^{\pm i\widehat{E}_{\bf
k}t'/\hbar}=\frac{1}{2}\sum_\alpha e^{\pm iE_{{\bf
k}\alpha}t'/\hbar}(1+\alpha\hat{\mathbf{\Omega}}_{\bf
k}\cdot\widehat{\mbox{\boldmath$\sigma$}}),
\end{equation}
where $\alpha=\pm 1$, $\hat{\mathbf{\Omega}}_{\bf
k}=\mathbf{\Omega}_{\bf k}/|\mathbf{\Omega}_{\bf k}|$, and
$E_{{\bf k}\pm}=\epsilon_{\bf
k}\pm\frac{\hbar}{2}|\mathbf{\Omega}_{\bf k}|$ are the
eigenvalues of $H_0$ [Eq.(\ref{ham1})]. At this point, averaging
over the impurity positions ${\bf R}_i$ restores the translational
invariance: $\langle\widehat{\rho}_{{\bf k}{\bf
k}'}(t)\rangle_{\rm imp}=\delta_{{\bf k},{\bf
k}'}\widehat{\overline{\rho}}_{\bf k}(t)$, and, finally, the
equation of motion of electron spins for a given wave vector
${\bf k}$, ${\bf S}_{\bf k}=\frac{\hbar}{2}{\rm
Tr}(\widehat{\mbox{\boldmath$\sigma$}}\widehat{\overline{\rho}}_{\bf
k})$, is obtained by integrating over $t'$. By retaining only the
scattering contributions (Boltzmann approximation)
\cite{lechner}, the final result is therefore:
\begin{widetext}
\begin{eqnarray}
\label{dm20} \frac{d{\bf S}_{\bf k}}{dt} &=& \mathbf{\Omega}_{\bf
k}\times {\bf S}_{\bf k}-\frac{2\pi n}{\hbar}\sum_{\bf
p}V^2_{{\bf p}{\bf
k}}\frac{1}{4}\sum_{\alpha\beta}\left\{[1-\alpha\beta(\hat{\mathbf{\Omega}}_{\bf
k}\cdot\hat{\mathbf{\Omega}}_{\bf p})]({\bf S}_{\bf k}-{\bf
S}_{\bf p})+\alpha\beta\hat{\mathbf{\Omega}}_{\bf
k}[\hat{\mathbf{\Omega}}_{\bf p}\cdot ({\bf S}_{\bf k}-{\bf
S}_{\bf p})]+\alpha\beta\hat{\mathbf{\Omega}}_{\bf
p}[\hat{\mathbf{\Omega}}_{\bf k}\cdot ({\bf S}_{\bf k}-{\bf
S}_{\bf p})]\right. \nonumber \\
& &\left.+\hbar(\alpha\hat{\mathbf{\Omega}}_{\bf
k}+\beta\hat{\mathbf{\Omega}}_{\bf p})(f_{\bf k}-f_{\bf
p})\right\}\delta(E_{{\bf k}\alpha}-E_{{\bf p}\beta}).
\end{eqnarray}
\end{widetext}
where $n$ is the impurity concentration and $f_{\bf
k}=\frac{1}{2}{\rm Tr}(\widehat{\overline{\rho}}_{\bf k})$ is the
electron occupation distribution function. Equation (\ref{dm20})
is very general, and holds true for both bulk and low-dimensional
systems with no restrictions on the particular form of
$\mathbf{\Omega}_{\bf k}$. In this letter, Eq.(\ref{dm20}) is
used to find the time evolution of the electron spin polarization
${\bf S}=\sum_{\bf k}{\bf S}_{\bf k}$ for a two-dimensional
electron gas confined in the $x-y$ plane and subjected to a
Rashba SO interaction. In such a case, the effective SO field is
$\mathbf{\Omega}_{\bf k}=(\gamma_R k_y,-\gamma_R k_x,0)$ and the
corresponding SO split bands depend solely on the magnitude $k$
of the wave vector and can be written as $E_{\bf
k\alpha}=E_{k\alpha}=\frac{\hbar^2}{2m}(k+\alpha k_R)^2$, where
$m$ is the electron mass and $k_R=\frac{m}{2\hbar}\gamma_R$
\cite{note}. In the following, instead of using $\Delta_{\rm
so}=\hbar\gamma_R k_F$, the SO splitting will be parametrized by
the Rashba energy
$\epsilon_R=\frac{\hbar^2k_R^2}{2m}=\frac{m}{8}\gamma_R^2$, that
is the minimum inter-band excitation energy for an electron
sitting at the bottom of the lower band [see Fig.\ref{fig1}(a)].

By assuming that the momentum dependence of the impurity potential
can be neglected, $V_{{\bf k}{\bf p}}=V$, ${\bf S}_{\bf k}$ is
conveniently decomposed in its even, ${\bf S}^e_{\bf k}$, and
odd, ${\bf S}^o_{\bf k}$, parts with respect to ${\bf k}$. It is
then clear that ${\bf S}=\sum_{\bf k}{\bf S}^e_{\bf k}$ and, from
Eq.(\ref{dm20}), $d{\bf S}/dt=\sum_{\bf k}\mathbf{\Omega}_{\bf
k}\times {\bf S}_{\bf k}=\sum_{\bf k}\mathbf{\Omega}_{\bf
k}\times {\bf S}^o_{\bf k}$, where $\mathbf{\Omega}_{-{\bf
k}}=-\mathbf{\Omega}_{\bf k}$ has been used. Of course, $d{\bf
S}/dt$ is also equal to $\sum_{\bf k}d{\bf S}^e_{\bf k}/dt$, so
that, after a further derivative with respect to time, the
equation of motion can be recast in the following form:
\begin{equation}
\label{spin1} \sum_{\bf k}\left(\frac{d^2{\bf S}^e_{\bf
k}}{dt^2}-\mathbf{\Omega}_{\bf k}\!\times\!\frac{d{\bf S}^o_{\bf
k}}{dt}\right)=0.
\end{equation}
By taking the vector product with $\mathbf{\Omega}_{\bf k}$, the
odd part of Eq.(\ref{dm20}) can be rewritten as:
\begin{eqnarray}
\label{dm21} \mathbf{\Omega}_{\bf k}\!\times\!\frac{d{\bf
S}^o_{\bf k}}{dt} &=&\mathbf{\Omega}_{\bf
k}\!\times\!(\mathbf{\Omega}_{\bf k}\!\times\!{\bf S}^e_{\bf
k})-\frac{2\pi nV^2}{4\hbar}\sum_{\alpha\beta}\sum_{{\bf
p}}\left[\mathbf{\Omega}_{\bf
k}\!\times{\bf S}^o_{\bf k}\right.\nonumber \\
& &+\alpha\beta\frac{k}{p}\mathbf{\Omega}_{\bf p}\!\times\!{\bf
S}^o_{\bf p}-\alpha\beta\frac{k}{p}\hat{\mathbf{\Omega}}_{\bf
k}\cdot(\mathbf{\Omega}_{\bf p}\!\times\!{\bf S}^o_{\bf
p})\,\hat{\mathbf{\Omega}}_{\bf k}]\nonumber \\
& &\times\delta(E_{k\alpha}-E_{p\beta}),
\end{eqnarray}
where the summation over momenta has cancelled all terms odd in
${\bf p}$ and the identity $(\hat{\mathbf{\Omega}}_{\bf
k}\cdot\hat{\mathbf{\Omega}}_{\bf p})(\mathbf{\Omega}_{\bf
k}\!\times\!{\bf S}^o_{\bf p})-(\mathbf{\Omega}_{\bf
k}\!\times\!\hat{\mathbf{\Omega}}_{\bf
p})(\hat{\mathbf{\Omega}}_{\bf k}\cdot{\bf S}^o_{\bf
p})=[\mathbf{\Omega}_{\bf p}\!\times\!{\bf S}^o_{\bf
p}-\hat{\mathbf{\Omega}}_{\bf k}\cdot(\mathbf{\Omega}_{\bf
p}\!\times\!{\bf S}^o_{\bf p})\,\hat{\mathbf{\Omega}}_{\bf
k}](k/p)$ has been used. From Eq.(\ref{dm20}), the equation of
motion of ${\bf S}^e_{\bf k}$ is instead given by:
\begin{eqnarray}
\label{dm22} \frac{d{\bf S}^e_{\bf k}}{dt}&=&\mathbf{\Omega}_{\bf
k}\!\times\!{\bf S}^o_{\bf k}-\frac{2\pi
nV^2}{4\hbar}\sum_{\alpha\beta}\sum_{\bf p}({\bf S}^e_{\bf k}-{\bf
S}^e_{\bf p} \nonumber \\
& &+\hbar\alpha\hat{\mathbf{\Omega}}_{\bf k}f^o_{\bf
k}-\hbar\beta\hat{\mathbf{\Omega}}_{\bf p}f^o_{\bf p})
\,\delta(E_{k\alpha}-E_{p\beta}),
\end{eqnarray}
where $f^o_{\bf k}=\frac{1}{2}(f_{\bf k}-f_{-{\bf k}})$ is the
odd part of $f_{\bf k}$.

By using Eqs.(\ref{dm21}) the term $\mathbf{\Omega}_{\bf
k}\!\times d{\bf S}^o_{\bf k}/dt$ in Eq.(\ref{spin1}) can be
eliminated in favor of ${\bf S}^o_{\bf k}$ and ${\bf S}^e_{\bf
k}$. Next, by using Eq.(\ref{dm22}), also the terms containing
${\bf S}^o_{\bf k}$ can be eliminated and Eq.(\ref{spin1})
reduces to an equation of motion of the component ${\bf S}^e_{\bf
k}$ only, that is sufficient to find the time evolution of ${\bf
S}=\sum_{\bf k}{\bf S}^e_{\bf k}$. Let us consider the
$z$-component of ${\bf S}$, $S_z$, for which the presence of
$f^o_{\bf k}$ in Eq.(\ref{dm22}) has not effect since these terms
have zero component in the $z$-direction. In this way,
Eq.(\ref{spin1}) reduces to:
\begin{equation}
\label{spin2} \int_0^\infty\!\frac{dk
k}{2\pi}\left[\frac{d^2S^e_{zk}}{dt^2}+\left(\gamma_R^2k^2+
\frac{\chi_k}{4\tau_{\rm
p}^2}\right)S^e_{zk}+\frac{\Gamma_k}{\tau_{\rm p}}\frac{d
S^e_{zk}}{dt}\right]=0,
\end{equation}
where $S^e_{zk}=\int_0^{2\pi}\!d\phi\, S^e_{z{\bf k}}/2\pi$, with
$\phi$ being the angle between the directions of ${\bf k}$ and the
$x$-axis, $\tau_{\rm p}^{-1}=\frac{2\pi}{\hbar}nV^2N_0$ is the
momentum relaxation rate for a two-dimensional electron gas with
zero SO splitting and density of states (DOS) $N_0=m/2\pi\hbar^2$,
and:
\begin{eqnarray}
\label{gamma1} \Gamma_k
\!\!&=&\!\!\frac{1}{4N_0}\sum_{\alpha\beta}\!
\int_0^\infty\!\frac{dpp}{2\pi}\left(1+\alpha\beta\frac{p}{k}\right)
\delta(E_{k\alpha}-E_{p\beta}), \\
\label{chi1} \chi_k\!\!&=&\!\!\frac{1}{N_0}\sum_{\alpha\beta}\!
\int_0^\infty\!\frac{dpp}{2\pi}(\Gamma_k-\Gamma_p)\delta(E_{k\alpha}-E_{p\beta}).
\end{eqnarray}
A solution to Eq.(\ref{spin2}) is obtained by equating to zero the
expression within square brackets, which leads to a homogeneous
differential equation of the second order for $S^e_{zk}(t)$. In
this way the functions $\Gamma_k$ and $\chi_k$ assume
respectively the meaning of renormalization of the damping term
and of shift of the (bare) precessional frequency $\gamma_R k$. It
can be easily realized from Eqs.(\ref{gamma1},\ref{chi1}) that in
the weak SO limit $\epsilon_F/\epsilon_R\rightarrow \infty$, for
which $E_{k\pm}\rightarrow \frac{\hbar^2k^2}{2m}$, both the
damping renormalization and the frequency shift are absent
($\Gamma_k=1$ and $\chi_k=0$), indicating that these quantities
stem from additional intra- and inter-band scattering channels
opened when $\epsilon_F/\epsilon_R$ is finite. Let us take a
closer look at $\Gamma_k$ and $\chi_k$ by performing the
integration over $p$ in Eqs.(\ref{gamma1},\ref{chi1}):
\begin{equation}
\label{gamma3} \Gamma_k = \left\{
\begin{array}{ll}
\frac{ k_R}{ k_R-k} & \,\,\,0\leq k\leq k_R \\
1+\frac{ (2k_R-k)k_R}{ (k-k_R)k} & \,\,\,k_R\leq k \leq 2k_R \\
1 & \,\,\,2k_R\leq k
\end{array}\right.
\end{equation}
\begin{equation}
\label{chi3} \chi_k = \left\{
\begin{array}{ll}
\frac{ 2k^2+3k_Rk}{ k_R^2-k^2} & \,\,\,0\leq k\leq k_R \\
\frac{ (2k_R-k)k}{ k^2-k_R^2} & \,\,\,k_R\leq k \leq 2k_R \\
-\frac{ (k-2k_R)^2}{ (k-k_R)(3k_R-k)} & \,\,\,2k_R\leq k \leq 3k_R \\
-\frac{ (4k_R-k)k_R}{ (k-k_R)(k-3k_R)} &
\,\,\,3k_R \leq k\leq 4k_R \\
0 & \,\,\,4k_R\leq k
\end{array}\right.
\end{equation}
For $k>4k_R$, $\Gamma_k$ and $\chi_k$ are the same as in the zero
SO limit, while for lower momenta they acquire a strong $k$
dependence (plotted in Fig.\ref{fig1}(c)) arising from the
combined effect of the reduced dimensionality ($D=2$) and the SO
interaction. In particular, the divergence of $\Gamma_k$ at
$k=k_R$ and those of $\chi_k$ at $k=k_R$ and $k=3k_R$ are due to
scattering processes probing the SO induced van Hove singularity
of the DOS of the lower sub-band which diverges as $N_-(E)\propto
\sqrt{\epsilon_R/E}$ [see Fig.\ref{fig1}(b)]. As it is shown
below, such  strong $k$ dependence has important consequences on
the spin polarization dynamics.

Let us now turn to evaluate the explicit time dependence of $S_z$.
From Eq.(\ref{spin2}), a general solution for $S^e_{zk}(t)$ is
given by a linear combination of $\exp\!\left( -\frac{
\Gamma_k\pm\sqrt{\Sigma_k}}{2\tau_{\rm p}} t \right)$, where
$\Sigma_k=\Gamma_k^2-\chi_k-(2\tau_{\rm p} \gamma_R k)^2$, whose
coefficients are fixed by imposing some initial conditions. If at
$t=0$ electrons have been prepared with a non-equilibrium
spin-state occupation but equilibrium distribution for each spin
branch then, at the lowest order in the initial weak spin
imbalance $\delta\mu=(\mu_\uparrow-\mu_\downarrow)/2$,
$S^e_{zk}(0)$ is simply:
\begin{equation}
\label{spin4} S^e_{zk}(0)=-\frac{\delta\mu}{\gamma_R
k}\sum_\alpha\alpha f_0(E_{k\alpha}-\mu) ,
\end{equation}
where $f_0(x)=(\exp(x/T)+1)^{-1}$ is the Fermi distribution
function, $T$ is the temperature and $\mu\geq 0$ is the chemical
potential. Furthermore, by imposing that
$\lim_{t\rightarrow\infty}|S_{zk}(t)| <\infty$ and by arbitrarily
choosing $dS_z(0)/dt=0$ for $\tau_{\rm p}^{-1}=0$, at zero
temperature ($\mu=\epsilon_F$) $S_z(t)$ is readily found to be:
\begin{eqnarray}
\label{spin3} S_z(t)&\!\!\!=\!\!\!\!\!&-\frac{\delta\mu}{2\pi
\gamma_R }\int_{k_<}^{k_>} \! dk\!\left[\theta(\Sigma_k
-\Gamma_k^2)
\exp\!\left(-\frac{\Gamma_k+\sqrt{\Sigma_k}}{2\tau_{\rm
p}}t\right)\right. \nonumber \\
& &\left. +\theta(\Gamma_k^2-\Sigma_k) \exp\!\left(-\frac{\Gamma_k
}{2\tau_{\rm p}}t\right)
\cosh\!\left(\frac{\sqrt{\Sigma_k}}{2\tau_{\rm p}}t\right)\right],
\end{eqnarray}
where $\theta$ is the unit step function, $k_\gtrless =k_R
(\sqrt{\epsilon_F/\epsilon_R}\pm 1)$ for $\epsilon_F/\epsilon_R >
1$ and $k_\gtrless= k_R (1\pm\sqrt{\epsilon_F/\epsilon_R})$ for
$\epsilon_F/\epsilon_R < 1$ [see Fig.\ref{fig1}(a)]. For
$\epsilon_F/\epsilon_R\rightarrow \infty$, $k_\gtrless\rightarrow
k_F$ and Eq.(\ref{spin3}) reduces to the classical formula
\begin{equation}\label{DP0}
S_z(t)=S_z(0)\exp\!\!\left(-\frac{t}{2\tau_p}\right)
\cosh\!\!\left(\frac{t\sqrt{1-(2\tau_p\Omega_R)^2}}
{2\tau_p}\right),
\end{equation}
where $S_z(0)=-\hbar\delta\mu N_0$ and $\Omega_R=\gamma_R k_F$ is
the Rashba frequency which characterizes the (damped) spin
precession behavior $S_z(t)\approx S_z(0)\exp(-\frac{t}{2\tau_{\rm
p}}) \cos(\Omega_R t)$ for $2\tau_p\Omega_R\gg 1$ and the DP
relaxational decay $S_z(t)\approx S_z(0)\exp(-\tau_{\rm
p}\Omega_R^2 t)$ for $2\tau_p\Omega_R\ll 1$ \cite{fabian}.

\begin{figure}[t]
\protect
\includegraphics[scale=0.34]{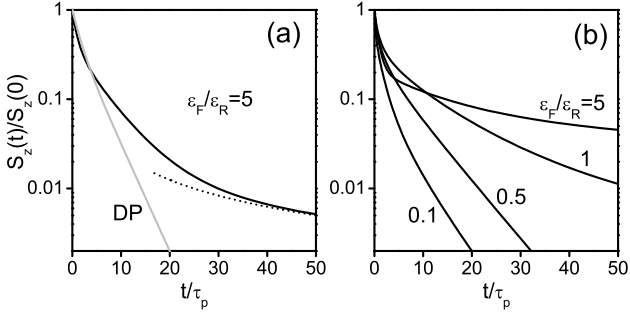}
\caption{(a): Spin polarization time evolution from
Eq.(\ref{spin3}) (solid line) for $\tau_p=3.3$ fs, $\epsilon_R=5$
meV and $\epsilon_F=25$ meV. The dotted line is the power law
decay of Eq.(\ref{spin7}) while the grey solid line is
Eq.(\ref{DP0}). (b): Crossover from the power law decay of
(\ref{spin7}) and the fast relaxation decay of (\ref{spin8})
obtained for $\tau_p=3.3$ fs, $\epsilon_F=5$ meV and several
values of $\epsilon_R$.} \label{fig2}
\end{figure}

For finite values of $\epsilon_F/\epsilon_R$, Eq.(\ref{spin3})
starts to deviate from the classical regime (\ref{DP0}).  Let us
first consider $\epsilon_F/\epsilon_R\geq 25$. In this case, the
integration over $k$ in Eq.(\ref{spin3}) spans values necessarily
larger than $4k_R$, where, according to
Eqs.(\ref{gamma3},\ref{chi3}), $\Gamma_k=1$ and $\chi_k=0$ [see
also Fig.\ref{fig1}(c)]. For weak impurity scattering
($2\tau_{\rm p}\gamma_Rk_< \gg 1$) $\Sigma_k$ is negative and
Eq.(\ref{spin3}) reduces to:
\begin{equation}
\label{spin5} S_z(t)\approx S_z(0)
\exp\left(\!-\frac{t}{2\tau_{\rm p}}\!\right) \frac{\sin(\Omega_>
t)-\sin(\Omega_< t)}{(\Omega_> -\Omega_<) t},
\end{equation}
where $\Omega_\gtrless=\gamma_Rk_\gtrless$. The main feature of
Eq.(\ref{spin5}) is that $S_z(t)$ oscillates with two different
Rashba frequencies $\Omega_\gtrless$ associated with the two SO
splitted Fermi surfaces, Fig.\ref{fig1}(a). Two distinct Rashba
frequencies characterize also the relaxation regime obtained when
the scattering with impurities is strong enough that $2\tau_{\rm
p}\gamma_Rk_> \ll 1$ (and so $\Sigma_k>0$) holds true. Also in
this case the integration of Eq.(\ref{spin3}) is elementary and
\begin{equation}
\label{spin6} S_z(t)\!\approx\!
S_z(0)\frac{\sqrt{\pi}}{2}\frac{{\rm
erf}\!\left(\!\sqrt{\tau_{\rm p}\Omega_>^2 t}\right)-{\rm
erf}\!\left(\!\sqrt{\tau_{\rm p}\Omega_<^2
t}\right)}{\sqrt{\tau_{\rm p}(\Omega_>-\Omega_<)^2 t}},
\end{equation}
where ${\rm erf}$ is the error function.

The spin precession and relaxation regimes of
Eqs.(\ref{spin5},\ref{spin6}) are governed solely by the enhanced
momentum phase space settled by finite values of
$\epsilon_F/\epsilon_R$. Instead, for $\epsilon_F/\epsilon_R <
25$, also the momentum dependence of $\Gamma_k$ and $\chi_k$
becomes relevant, leading to important anomalous features of the
spin dynamics. One of these is particularly striking and it is
found when $1\leq\epsilon_F/\epsilon_R\leq 9$. In this case, the
integration over $k$ in Eq.(\ref{spin3}) crosses the point
$k=2k_R$ where $\chi_k$ changes sign [see Fig.\ref{fig1}(c)].
Hence, if $4\tau_{\rm p}\gamma_Rk_R \ll 1$, $\sqrt{\Sigma_k}$ can
be expanded as $\Gamma_k-(2k_R-k)/3k_R$ for $k<2k_R$, while when
$k>2k_R$  $\chi_k$ becomes negative leading to exponentially
small contributions to $S_z(t)$ for sufficiently long times.
Eq.(\ref{spin3}) then can be approximated to:
\begin{equation}
\label{spin7} S_z(t)\approx -\frac{\delta\mu}{2\pi \gamma_R}
\int_{k_<}^{2k_R}\!\frac{dk}{2}\,e^{-\frac{(2k_R-k)}{6\tau_{\rm
p}k_R}t}\approx \frac{3S_z(0)}{2}\frac{\tau_{\rm p}}{t}.
\end{equation}
The surprising result of Eq.(\ref{spin7}) provides the rather
interesting prediction that, for sufficiently strong SO
interaction and momentum scattering, the spin polarization decays
as a power law rather than exponentially. In this case therefore
the memory of the initial spin polarization can be much longer
lived than in the DP regime, as shown in Fig. \ref{fig2}(a).
Another striking feature is that obtained in the extreme
$\epsilon_F/\epsilon_R \ll 1$ limit in which the integration of
Eq.(\ref{spin3}) becomes restricted to a narrow region around
$k=k_R$ where both $\Gamma_k$ and $\chi_k$ diverge as
$1/|k-k_R|$, so that Eq.(\ref{spin3}) becomes:
\begin{equation}
\label{spin8} S_z(t)\propto\exp\!\left(-\frac{t}{8\tau_{\rm p}
}\right),
\end{equation}
indicating that for extremely strong SO interaction, momentum
scattering \emph{increases} the spin polarization decay. The
power decay of (\ref{spin7}) and the fast relaxation regime of
(\ref{spin8}) are plotted in Fig.\ref{fig2}(b) from a numerical
integration of Eq.(\ref{spin3}) for $\tau_p=3.3$ fs,
$\epsilon_F=5$ meV and $\epsilon_R$ ranging from $50$ meV down to
$1$ meV.

To conclude, the kinetic equations describing the time evolution
of the spin polarization have been formulated for arbitrary
strength of the SO interaction. Explicit solutions for quantum
wells with Rashba-like SO interactions predict the failure of the
DP relaxation formula for sufficiently strong SO couplings. In
particular, the memory of the initial spin state can be strongly
enhanced or reduced depending on $\epsilon_F/\epsilon_R$,
suggesting an alternative route for spin manipulation in
spintronic applications.

I thank E. Cappelluti for fruitful discussions.

\end{document}